\documentclass[10pt,A4paper,conference]{IEEEtran}
\usepackage{amsmath,color}
\newcommand{\F}{\mathbf{F}}
\newcommand{\Z}{\mathbf{Z}}

\newcommand{\B}{\mathcal{BCH}}
\newcommand{\ceil}[1]{\left\lceil #1\right\rceil}

\newcommand{\hdual}{{\bot_h}}

\newtheorem{theorem}{Theorem}
\newtheorem{lemma}[theorem]{Lemma}
\newtheorem{corollary}[theorem]{Corollary}
\newtheorem{example}[theorem]{Example}

\begin{document}
% paper title
\title{Primitive Quantum BCH Codes over Finite Fields}
% author names and affiliations
% use a multiple column layout for up to three different
% affiliations
\author{\authorblockN{Salah A. Aly, Andreas Klappenecker, Pradeep
Kiran Sarvepalli}
\authorblockA{Department  of Computer Science\\
Texas A\&M University\\
College Station, TX 77843-3112, USA \\
Email: \{salah,klappi,pradeep\}@cs.tamu.edu}   }
\maketitle

\begin{abstract}
An attractive feature of BCH codes is that one can infer valuable
information from their design parameters (length, size of the finite
field, and designed distance), such as bounds on the minimum distance 
and dimension of the code. In this paper, it is shown that one can
also deduce from the design parameters whether or not a primitive,
narrow-sense BCH contains its Euclidean or Hermitian dual code. This
information is invaluable in the construction of quantum BCH codes. A
new proof is provided for the dimension of BCH codes with small
designed distance, and simple bounds on the minimum distance of such
codes and their duals are derived as a consequence. These results
allow us to derive the parameters of two families of primitive quantum
BCH codes as a function of their design parameters.
\end{abstract}

\section{Introduction}
Let $\alpha$ denote a primitive element in the finite field
$\F_{q^m}$. We set $n=q^m-1$ and denote by $\delta$ an integer in the
range $2\le \delta \le n$. Recall that a cyclic code of length $n$
over $\F_q$ is called a primitive, narrow-sense BCH code with designed
distance~$\delta$ if its generator polynomial is of the form
$$g(x)=\prod_{z\in Z} (x-\alpha^z)\quad \mbox{ with} \quad Z=C_1\cup
\cdots \cup C_{\delta-1},$$ where $C_x=\{ xq^k\bmod n \,|\, 0\le k<
m\,\}$ denotes the $q$-ary cyclotomic coset of $x$ modulo~$n$.  
We refer to such a code as a $\B(n,q;\delta)$ code, and call 
$Z$ the defining set of the code. The basic properties of
these classical codes are discussed, for example, in the books
\cite{huffman03,kabatiansky05,macwilliams77}.

Given a classical BCH code, we can use one of the following well-known
constructions to derive a quantum stabilizer code:
\begin{enumerate}
\item If there exists a classical linear $[n,k,d]_q$ code $C$ such
that $C^\perp\subseteq C$, then there exists an $[[n,2k-n,\ge d]]_q$
stabilizer code that is pure to $d$. If the minimum distance of
$C^\perp$ exceeds $d$, then the quantum code is pure and has minimum
distance $d$.
\item If there exists a classical linear $[n,k,d]_{q^2}$ code $D$ such
that $D^\hdual\subseteq D$, then there exists an $[[n,2k-n,\ge
d]]_{q}$ stabilizer code that is pure to $d$. If the minimum
distance of $D^\hdual$ exceeds $d$, then the quantum code is pure and
has minimum distance $d$.
\end{enumerate}
The orthogonality relations are defined in the \textit{Notations}\/ at
the end of this section.  Examples of certain binary quantum BCH codes
have been given in \cite{calderbank98,grassl99b,grassl04,steane96}.

Our goal is to derive the parameters of the quantum stabilizer code as
a function of their design parameters $n$, $q$, and $\delta$ of the
associated primitive, narrow-sense BCH code $C$. This entails the
following tasks:
\begin{enumerate}
\item[a)] Determine the design parameters for which $C^\perp\subseteq C$;
\item[b)] determine the dimension of $C$; 
\item[c)] bound the minimum weight in $C\setminus C^\perp$. 
\end{enumerate}
In case $q$ is a perfect square, we would also like to answer the
Hermitian versions of questions a) and c):
\begin{enumerate}
\item[a')] Determine the design parameters for which $C^\hdual\subseteq C$;
\item[c')] bound the minimum weight in $C\setminus C^\hdual$.
\end{enumerate}
\smallskip

To put our work into perspective, we sketch our results and give a
brief overview of related work. 

Let $C$ be a primitive, narrow-sense BCH code $C$ of length $n=q^m-1$,
$m\ge 2$, over $\F_q$ with designed distance $\delta$.

To answer question a), we prove in Theorem~\ref{th:dual} that
$C^\perp\subseteq C$ holds if and only if $\delta\le
q^{\ceil{m/2}}-1-(q-2)[m \text{ odd}]$. The significance of this
result is that allows one to identify all BCH codes that can be used
in the quantum code construction~1). Fortunately, this question can be
answered now without computations. Steane proved in~\cite{steane99}
the special case $q=2$, which is easier to show, since in this case
there is no difference between even and odd $m$.

In Theorem~\ref{th:hdual}, we answer question a') and show that
$C^\hdual\subseteq C$ if and only if $\delta\le q^{(m+[\text{$m$
even}])/2}-1-(q-2)[\text{$m$ even}]$, where we assume that
$q$ is a perfect square.  This result allows us to determine all
primitive, narrow-sense BCH codes that can be used in construction
2). We are not aware of any prior work concerning the Hermitian case.

In the binary case, an answer to question b) was given by MacWilliams
and Sloane~\cite[Chapter 9, Corollary 8]{macwilliams77}.  Apparently,
Yue and Hu answered question b) in the case of small designed
distances~\cite{yue96}. We give a new proof of this result in
Theorem~\ref{th:bchdimension} and show that the dimension
$k=n-m\ceil{(\delta-1)(1-1/q)}$ for $\delta$ in the range $2\le
\delta<q^{\ceil{m/2}}+1$. As a consequence of our answer to b), we
obtain the dimensions of the quantum codes in constructions 1) and 2).

Finding the true minimum distance of BCH codes is an open problem for
which a complete answer seems out of reach, see~\cite{charpin98}.  As
a simple consequence of our answer to b), we obtain better bounds on
the minimum distance for some BCH codes, and we derive simple bounds
on the (Hermitian) dual distance of BCH codes with small designed
distance, which partly answers c) and c').

In Section~\ref{sec:qcodes}, all these results are used to derive two
families of quantum BCH codes. Impatient readers should now browse
this section to get the bigger picture. Theorem~\ref{sh:euclid} yields
the result that one obtains using construction 1). 
The result of construction 2) is given in
Theorem~\ref{sh:hermite}.

\medskip
\textit{Notations.} We denote the ring of integers by $\mathbf{Z}$ and
a finite field with $q$ elements by $\mathbf{F}_q$. We follow Knuth
and attribute to $[P(k)]$ the value 1 if the property $P(k)$ of the
integer $k$ is true, and 0 otherwise. For instance, we have $[k \text{
even}]= k-1\bmod 2$, but the left hand side seems more readable.  If
$x$ and $y$ are vectors in $\F_q^n$, then we write $x\perp y$ if and
only if $x\cdot y=0$. Similarly, if $x$ and $y$ are vectors in
$\F_{q^2}^n$, then we write $x\,\hdual\, y$ if and only if $x^q\cdot
y=0$.  

%%%%%%%%%%%%%%%%%%%%%%%%%%%%%%%%%%%%%%%%%%%%%%%%%%%%%%%%%%%%%%%%
%%% Selforthogonality %%%%%%%%%%%%%%%%%%%%%%%%%%%%%%%%%%%%%%%%%%
%%%%%%%%%%%%%%%%%%%%%%%%%%%%%%%%%%%%%%%%%%%%%%%%%%%%%%%%%%%%%%%%
\section{Euclidean Dual Codes}
Recall that the Euclidean dual code $C^\perp$ of a code $C\subseteq
\F_q^n$ is given by $C^\perp = \{ y\in \F_q^n \,|\,
x\cdot y =0 \mbox{ for all } x \in C \}.$ Steane showed
in~\cite{steane99} that a primitive binary BCH code of length $2^m-1$
contains its dual if and only if its designed distance $\delta$
satisfies $\delta \leq 2^{\lceil m/2\rceil}-1$. In this section we derive 
a similar condition for nonbinary BCH codes.

\begin{lemma}\label{th:selforthogonal}
Suppose that $\gcd(n,q)=1$. A cyclic code of length $n$ over $\F_q$ with
defining set $Z$ contains its Euclidean dual code if and only if $Z\cap
Z^{-1} = \emptyset$, where $Z^{-1}$ denotes the set 
$Z^{-1}=\{-z\bmod n\mid z \in Z \}$.
\end{lemma}
\begin{proof}
See, for instance,~\cite[Lemma~2]{grassl97} or~\cite[Theorem
4.4.11]{huffman03}.
\end{proof}

\begin{theorem} \label{th:dual}
A primitive, narrow-sense BCH code of length $q^m-1$, with $m\ge 2$,
over the finite field\/ $\F_q$ contains its dual code if and only if its
designed distance $\delta$ satisfies
$$\delta \leq
\delta_{\max}=q^{\lceil m/2\rceil}-1-(q-2)[m \textup{ odd}].$$
\end{theorem}
\begin{proof}
Let $n=q^m-1$.  The defining set $Z$ of a primitive, narrow-sense BCH
code $C$ of designed distance $\delta$ is given by $Z=C_1\cup
C_2\cdots\cup C_{\delta-1}$, where
$C_x=\{xq^j\bmod n \mid j\in \mathbf{Z} \}$.
\begin{enumerate}
\item We will show that the code
$C$ cannot contain its dual code if the designed distance
$\delta>\delta_{\max}$. Seeking a contradiction, we assume that the
defining set $Z$ contains the set $\{1,\dots,s\}$, where
$s=\delta_{\max}$. By Lemma~\ref{th:selforthogonal}, it suffices to
show that $Z\cap Z^{-1}$ is not empty.
If $m$ is even, then $s = q^{m/2}-1$, and $Z^{-1}$ contains the
element $-s q^{m/2} \equiv q^{m/2}-1 \equiv s \bmod n$, which means
that $Z\cap Z^{-1} \neq \emptyset$; contradiction. If $m$ is odd, then
$s=q^{(m+1)/2}-q+1$, and the element given by $-sq^{(m-1)/2}\equiv
q^{(m+1)/2}-q^{(m-1)/2} -1 \bmod n$ is contained in $Z^{-1}$. Since
this element is less than $s$ for $m\geq 3$, it is contained in $Z$,
so $Z\cap Z^{-1}\neq \emptyset$; contradiction.  Combining these two
cases, we can conclude that $\delta\leq q^{\lceil
m/2\rceil}-1-(q-2)[m\mbox{ is odd}]$ for $m\geq 2$.

\item For the converse, we prove that if $\delta\leq \delta_{\max}$,
then $Z\cap Z^{-1}=\emptyset$, which implies $C^\perp \subseteq C$ by
Lemma~\ref{th:selforthogonal}. It suffices to show that $\min C_{-x}
\geq \delta_{\max}$ for any coset $C_x$ in~$Z$. Since $1\leq
x<\delta_{\max}\leq q^{\lceil m/2\rceil}-1$, we can write $x$ as a
$q$-ary integer of the form $x=x_0+x_1q+\cdots+x_{m-1}q^{m-1}$ with
$0\le x_i<q$, and $x_i=0$ for $i\ge \lceil m/2\rceil.$ If
$\bar{y}=n-x$, then $\bar{y}= \bar{y}_0+\bar{y}_1q + \cdots +
\bar{y}_{m-1}q^{m-1} =\sum_{i=0}^{m-1}(q-1-x_i)q^i. $ Set $y=\min
C_{-x}$. We note that $y$ is a conjugate of $\bar{y}$.  Thus, the
digits of $y$ are obtained by cyclically shifting the digits of
$\bar{y}$.

\item[3a)] First we consider the case when $m$ is even. Then the $q$-ary
expansion of $x$ has at least $m/2$ zero digits. Therefore, at least
$m/2$ of the $\bar{y}_i$ are equal to $q-1$.  Thus, $y\geq
\sum_{i=0}^{m/2-1}(q-1) q^i=q^{m/2}-1=\delta_{\max}$.

\item[3b)] If $m$ is odd, then as $1\leq x<q^{(m+1)/2}-q+1$, we have
$m>1$ and $\bar{y}=\bar{y}_0+\bar{y}_1q + \cdots +
(\bar{y}_{(m-1)/2})q^{(m-1)/2} + (q-1)q^{(m+1)/2}+\cdots +
(q-1)q^{m-1}$. For $0\leq j\leq (m-1)/2$, we observe that $xq^j<n$,
and since $\bar{y}q^j \equiv -xq^j \bmod n, \bar{y}q^j  = n-xq^j
\geq q^m-1-(q^{(m+1)/2}-q)q^{(m-1)/2} =
q^{(m+1)/2}-1 \geq \delta_{\max}$. For $(m+1)/2 \leq j\leq m-1$, we
find that
\begin{eqnarray*}
\begin{split}
\bar{y}q^j \bmod n
&=\bar{y}_{m-j}+\cdots+\bar{y}_{(m-1)/2}q^{j-(m+1)/2} 
\\& + (q-1)q^{j-(m-1)/2} + \cdots + (q-1)q^{j-1}\\&+\bar{y}_0q^{j} + \cdots +
\bar{y}_{m-j-1}q^{m-1},\\
&\geq  (q^{(m-1)/2}-1)q^{j-(m-1)/2} +\bar{y}_0+\cdots\\&+\bar{y}_{(m-1)/2},\\
& \geq q^{(m+1)/2}-q+ 1 = \delta_{\max},
\end{split}
\end{eqnarray*}
where $\bar{y}_0+\cdots+\bar{y}_{(m-1)/2}\geq 1$ because
$x<q^{(m+1)/2}-q+1$. Hence $y =\min\{\bar{y}q^j\mid j \in
\mathbf{Z}\} \geq \delta_{\max}$ when $m$ is odd.
\end{enumerate}
Therefore a primitive BCH code contains its dual if and only if
$\delta \leq \delta_{\max}$, for $m\geq 2$.
\end{proof}

%%%%%%%%%%%%%%%%%%%%%%%%%%%%%%%%%%%%%%%%%%%%%%%%%%%%%%%%%%%%%%%%
%%% Hermitian Selforthogonality %%%%%%%%%%%%%%%%%%%%%%%%%%%%%%%%
%%%%%%%%%%%%%%%%%%%%%%%%%%%%%%%%%%%%%%%%%%%%%%%%%%%%%%%%%%%%%%%%
\section{Hermitian Dual Codes}
If the cardinality of the field is a perfect square, then we can
define another type of orthogonality relation for codes.  Recall that
if the code $C$ is a subspace of the vector space $\F_{q^2}^n$, then
its Hermitian dual code $C^{\perp_h}$ is given by $C^{\perp_h}=\{ y\in
\F_{q^2}^n\,|\, y^q\cdot x = 0 \mbox{ for all } x \in C\}$, where
$y^q=(y_1^q,\dots,y_n^q)$ denotes the conjugate of the vector
$y=(y_1,\dots,y_n)$. The goal of this section is to establish when a
primitive, narrow-sense BCH code contains its Hermitian dual code.

\begin{lemma}\label{th:hermitian}
Assume that $\gcd(n,q)=1$. A cyclic code of length $n$ over
$\F_{q^2}$ with defining set $Z$ contains its Hermitian dual code if
and only if $Z\cap Z^{-q} = \emptyset$, where $Z^{-q}=\{-qz \bmod n
\mid z \in Z \}$.
\end{lemma}
\begin{proof}
Let $N=\{0,1,\dots,n-1\}$. If $g(z)=\prod_{x\in Z} (z-\alpha^x)$ is
the generator polynomial of a cyclic code $C$, then
$h^\dagger(z)=\prod_{x\in N\setminus Z} (z-\alpha^{-qx})$ is the
generator polynomial of $C^{\perp_h}$.  Thus, $C^{\perp_h}\subseteq C$
if and only if $g(z)$ divides $h^\dagger(z)$. The latter condition is
equivalent to $Z\subseteq \{ -qx\,|\, x\in N\setminus Z\}$, which can
also be expressed as $Z\cap Z^{-q}=\emptyset$.
\end{proof}

\begin{theorem}\label{th:hdual}
A primitive, narrow-sense BCH code of length $q^{2m}-1$ over
$\F_{q^2}$, where $m\neq 2$, contains its Hermitian dual code if and
only if its designed distance $\delta$ satisfies
\begin{equation*}
\delta\leq \delta_{\max}=q^{m+[m \text{ even}]}-1-(q^2-2)[m \text{ even}].
\end{equation*}
\end{theorem}
\begin{proof}
Let $n=q^{2m}-1$. Recall that the defining set $Z$ of a primitive,
narrow-sense BCH code $C$ over the finite field $\F_{q^2}$ with
designed distance $\delta$ is given by $Z=C_1\cup \cdots \cup
C_{\delta-1}$ with $C_x=\{ xq^{2j}\bmod n\,|\, j\in \Z \}$.
\begin{enumerate}
\item We will show that the code $C$ cannot contain its Hermitian dual
code if the designed distance $\delta> \delta_{\max}$. Seeking a
contradiction, we assume that the defining set $Z$ contains
$\{1,\dots,s\}$, where $s=\delta_{\max}$. By Lemma~\ref{th:hermitian},
it suffices to show that $Z\cap Z^{-q}$ is not empty.
If $m$ is odd, then $s=q^{m}-1$. Notice that $n-qs
q^{2(m-1)/2}=q^{m}-1=s$, which means that $s\in Z\cap Z^{-q}$, and
this contradicts our assumption that this set is empty.
If $m$ is even, then $s=q^{m+1}-q^2+1$. We note that $n-qsq^{m-2} =
q^{m+1}-q^{m-1}-1 < s=q^{m+1}-q^2+1$, for $m>2$. It follows that $q^{m+1}-q^{m-1}-1\in Z\cap
Z^{-q}$, contradicting our assumption that this set is empty.
Combining the two cases, we can conclude that $s$ must be smaller than
the value $q^{m+[m \text{ even}]}-1-(q^2-2)[m \text{ even}]$.

\item For the converse, we show that if $\delta<\delta_{\max}$, then
$Z\cap Z^{-q}=\emptyset$, which implies $C^{\perp_h}\subseteq C$
thanks to Lemma~\ref{th:hermitian}. It suffices to show that
$\min \{n-qC_x\} \geq \delta_{\max}$ or, equivalently, that $\max
qC_x \leq n-\delta_{\max}$ holds for $1\le x\le \delta-1$.

\item If $m$ is odd, then the $q$-ary expansion of $x$ is of the form
$x=x_0+x_1q+\cdots+x_{m-1}q^{m-1}$, with $x_i=0$, for $m \leq i\leq
2m-1$ as $x<q^{m}-1$. So at least $m$ of the $x_i$ are equal to
zero, which implies $\max qC_x <q^{2m}-1-(q^m-1)=n-\delta_{\max}$.

\item Let $m$ be even and $qxq^{2j}$ be the $q^2$-ary conjugates of
$qx$. Since $x<q^{m+1}-q^2+1$, $x=x_0+x_1q+\cdots+x_{m}q^m$ and at
least one of the $x_i \leq q-2$. If $0\leq 2j\leq m-2$, then
$qxq^{2j}\leq q(q^{m+1}-q^2)q^{m-2} = q^{2m}-q^{m+1} = n-q^{m+1}+1<
n-\delta_{\max}$. If $2j=m$, then $qxq^{m} =
x_{m-1}+x_{m}q+0.q^2+\cdots+
0.q^{m}+x_0q^{m+1}\cdots+x_{m-2}q^{2m-1}$. We note that there occurs
a consecutive string of $m-1$ zeros and because one of the $x_i\leq
q-2$, we have $qxq^{2j}<n- q^{2}(q^{m-1}-1)-1\leq n-\delta_{\max}$.
For $m+2 \leq 2j\leq 2m-2$, we see that $qxq^{2j}<
n-q^4(q^{m-1}-1)<n-\delta_{\max}$.
\end{enumerate}
Thus we can conclude that the primitive BCH codes contain their
Hermitian duals when $\delta\leq q^{m+[m \text{ even}]}-1-(q^2-2)[m
\text{ even}]$.
\end{proof}

%%%%%%%%%%%%%%%%%%%%%%%%%%%%%%%%%%%%%%%%%%%%%%%%%%%%%%%%%%%%%%%%
%%% Dimension and Minimum Distance %%%%%%%%%%%%%%%%%%%%%%%%%%%%%
%%%%%%%%%%%%%%%%%%%%%%%%%%%%%%%%%%%%%%%%%%%%%%%%%%%%%%%%%%%%%%%%
\section{Dimension and Minimum Distance}
In this section we determine the dimension of primitive, narrow-sense
BCH codes of length $n$ with small designed distance. Furthermore, we
derive bounds on the minimum distance of such codes and their duals.

\subsection{Dimension}
First, we make some simple observations about cyclotomic cosets that
are essential in our proof.

\begin{lemma} \label{th:bchcosetsize}
If\/ $q$ be a power of a prime, $m$ a positive integer and $n=q^m-1$,
then all $q$-ary cyclotomic cosets
$C_x=\{xq^\ell\bmod n\,|\,\ell \in \Z\}$ with $x$ in the range
$1\le x< q^{\lceil m/2\rceil}+1$
have cardinality $|C_x|=m$.
\end{lemma}
\begin{proof}
Seeking a contradiction, we assume that $|C_x|< m$. If $m=1$, then
$C_x$ would have to be the empty set, which is impossible.
If $m>1$, then $|C_x|<m$ implies that there must exist an integer
$j$\/ in the range $1\le j<m$ such that $j$ divides $m$ and
$xq^j\equiv x \mod n$. In other words, $q^m-1$ divides $x(q^j-1)$;
hence, $x\geq (q^m-1)/(q^j-1)$.

If $m$ is even, then $j\leq m/2$; thus, $x\geq q^{ m/2}+1$. If $m$ is
odd, then $j\leq m/3$ and it follows that $x\geq (q^m-1)/(q^{m/3}-1)$,
and it is easy to see that the latter term is larger than $q^{\lceil
m/2\rceil}+1$. In both cases this contradicts our assumption that
$1\le x\le q^{\lceil m/2\rceil}$; hence $|C_x|=m$.
\end{proof}

\begin{lemma}\label{th:disjointcosets}
Let $q$ be a power of a prime, $m$ a positive integer, and $n=q^m-1$.
Let $x$ and $y$ be integers in the range $1\le x,y< q^{\lceil
m/2\rceil}+1$ such that $x,y\not \equiv 0 \bmod q$. If $x \neq y$, then
the $q$-ary cosets of\/ $x$ and $y$ modulo~$n$
are disjoint, i.e., $C_x\neq C_y$. 
\end{lemma}

\begin{proof}
Seeking a contradiction, we assume that $C_x=C_y$. This assumption
implies that $y\equiv xq^\ell \bmod n$ for some integer $\ell$ in the range
$1\le \ell <m$.

If $xq^\ell<n$, then $xq^\ell \equiv 0\bmod q$; this contradicts our
assumption $y\not\equiv 0 \bmod q$, so we must have $xq^\ell \ge n$. 
It follows from the range of $x$ that $\ell$ must be at least 
${\lfloor m/2\rfloor}$. 

If $\ell={\lfloor m/2\rfloor}$, then we cannot find an admissible $x$
within the given range such that $y\equiv xq^{\lfloor m/2\rfloor}\bmod n$. 
Indeed, it follows from the inequality $xq^{\lfloor m/2\rfloor}\ge n$ 
that $x\ge q^{\lceil m/2\rceil}$, so $x$ must equal 
$q^{\lceil m/2\rceil}$, but that contradicts $x\not\equiv 0\bmod q$. 
Therefore, $\ell$ must exceed $\lfloor m/2\rfloor$. 

Let us write $x$ as a $q$-ary number
$x=x_0+x_1q+\cdots+x_{m-1}q^{m-1}$, with $0\le x_i<q$. Note that
$x_0\neq 0$ because $x\not\equiv 0\bmod q$. If $\lfloor
m/2\rfloor<\ell<m$, then $xq^\ell$ is congruent to $
y_0=x_{m-\ell}+\cdots + x_{m-1}q^{\ell-1}+x_0q^{\ell}+\cdots +
x_{m-\ell-1}q^{m-1}$ modulo $n$. We observe that $y_0\ge x_0q^\ell\ge
q^{\lceil m/2\rceil}$. Since $y\not\equiv 0 \bmod q$, it follows that 
$y=y_0\geq q^{\lceil m/2\rceil}+1$, contradicting the assumed range of $y$. 
\end{proof}

The previous two observations about cyclotomic cosets allow us to derive
a closed form for the dimension of a primitive BCH code. This result 
generalizes binary case \cite[Corollary~9.8, page~263]{macwilliams77}.
See also \cite{stichtenoth90} which gives estimates on the dimension of
BCH codes among other things. 
\begin{theorem}\label{th:bchdimension}
A primitive, narrow-sense BCH code of length $q^m-1$ over $\F_q$ with
designed distance $\delta$ in the range $2 \leq \delta \le q^{\lceil
m/2 \rceil}+1$ has dimension
\begin{equation}\label{eq:dimension}
k=q^m-1-m\lceil (\delta-1)(1-1/q)\rceil.
\end{equation}
\end{theorem}

\begin{proof}
The defining set of the code is of the form $Z=C_1\cup C_2\cdots
\cup C_{\delta-1}$, a union of at most $\delta -1$ consecutive 
cyclotomic cosets.
However, when $1\leq x\leq \delta-1$ is a multiple of $q$, then
$C_{x/q}=C_x$. Therefore, the number of cosets is reduced by
$\lfloor(\delta-1)/q \rfloor$. By Lemma~\ref{th:disjointcosets}, if
$x, y\not\equiv 0 \bmod q$ and $x\neq y$, then the cosets $C_x$ and
$C_y$ are disjoint. Thus, $Z$ is the union of $(\delta-1)-\lfloor
(\delta-1)/q\rfloor= \lceil (\delta-1)(1-1/q)\rceil$ distinct
cyclotomic cosets. By Lemma~\ref{th:bchcosetsize} all these cosets
have cardinality~$m$.  Therefore, the degree of the generator
polynomial is $m\lceil (\delta-1)(1-1/q)\rceil$, which proves
our claim about the dimension of the code.
\end{proof}

If we exceed the range of the designed distance in the hypothesis of
the previous theorem, then our dimension formula (\ref{eq:dimension})
is no longer valid, as our next example illustrates.

\begin{example}
Consider a primitive, narrow-sense BCH code of length $n=4^2-1=15$
over $\F_4$.  If we choose the designed distance $\delta =6 > 4^1+1$,
then the resulting code has dimension $k=8$, because the defining set $Z$ is given by 
$$ Z= C_1\cup C_2\cup \cdots \cup C_5 = \{1,4\}\cup \{2,8\}\cup
\{3,12\}\cup \{5\}.$$ The dimension formula (\ref{eq:dimension})
yields $4^2-1-2\lceil (6-1)(1-1/4)\rceil=7$, so the formula does
not extend beyond the range of designed distances given in
Theorem \ref{th:bchdimension}.
\end{example}

%%%%%%%%%%%%%%%%%%%%%%%%%%%%%%%%%%%%%%%%%%%%%%%%%%%%%%%%%%%%%%%%
%%% Minimum Distance %%%%%%%%%%%%%%%%%%%%%%%%%%%%%%%%%%%%%%%%%%%
%%%%%%%%%%%%%%%%%%%%%%%%%%%%%%%%%%%%%%%%%%%%%%%%%%%%%%%%%%%%%%%%
\subsection{Distance Bounds}

The true minimum distance $d_{min}$ of a primitive BCH code over
$\F_q$ with designed distance $\delta$ is bounded by $\delta\le
d_{min}\le q\delta-1$, see \cite[p.~261]{macwilliams77}.  If we apply
the Farr bound (essentially the sphere packing bound) using the
dimension given in Theorem~\ref{th:bchdimension}, then we obtain:

\begin{corollary}\label{th:mindist}
If $C$ is primitive, narrow-sense BCH code of length $q^m-1$ over
$\F_q$ with designed distance $\delta$ in the range $2\le \delta\le
q^{\lceil m/2\rceil}+1$ such that
\begin{eqnarray}\label{eqa3}
\sum_{i=0}^{\lfloor (\delta+1)/2\rfloor}
\binom{q^m-1}{i}
(q-1)^i >q^{m\lceil (\delta-1)(1-1/q)\rceil},
\end{eqnarray}
then $C$ has minimum distance $d= \delta$ or $\delta+1$; if,
furthermore, $\delta\equiv 0\bmod q$, then $d=\delta+1$.
\end{corollary}
\begin{proof}
Seeking a contradiction, we assume that the minimum distance~$d$ of
the code satisfies $d \geq \delta+2$. We know from
Theorem~\ref{th:bchdimension} that the dimension of the code is
$k=q^m-1-m\lceil (\delta-1)(1-1/q)\rceil.$ If we substitute this value of
$k$ into the sphere-packing bound
$$ q^{k}   \sum_{i=0}^{\lfloor (d-1)/2\rfloor}
\binom{q^m-1}{i}
(q-1)^i \leq q^n,
$$
then we obtain
\begin{eqnarray*}
\begin{split}
\sum_{i=0}^{\lfloor (\delta+1)/2\rfloor}\binom{q^m-1}{i}(q-1)^i &\le
\sum_{i=0}^{\lfloor (d-1)/2\rfloor}\binom{q^m-1}{i}(q-1)^i\\&\le
q^{m\lceil (\delta-1)(1-1/q)\rceil},
\end{split}
\end{eqnarray*}
but this contradicts condition~(\ref{eqa3}); hence,
$\delta\le d\le \delta+1$.

If $\delta\equiv 0\bmod q$, then the cyclotomic coset $C_\delta$ is
contained in the defining set $Z$ of the code because
$C_\delta=C_{\delta/q}$. Thus, the BCH bound implies that the minimum
distance must be at least $\delta+1$. 
\end{proof}

\begin{lemma}\label{th:dualdist} 
Suppose that $C$ is a primitive, narrow-sense BCH code of length
$n=q^m-1$ over $\F_q$ with designed distance $2\leq \delta\le
\delta_{\max}=q^{\lceil m/2\rceil}-1-(q-2)[m \textup{
odd}])$, then the dual distance $d^\perp \geq
\delta_{\max} + 1$.
\end{lemma}
\begin{proof}
Let $N=\{0,1,\ldots,n-1 \}$ and $Z_{\delta}$ be the defining set of
$C$. We know that $Z_{\delta_{\max}}\supseteq Z_{\delta}\supset
\{1,\ldots,\delta-1 \}$.  Therefore $N\setminus Z_{\delta_{\max}}
\subseteq N\setminus Z_{\delta}$.  Further, we know that $Z\cap
Z^{-1}=\emptyset$ if $2\leq \delta\leq \delta_{\max }$ from
Lemma~\ref{th:selforthogonal} and
Theorem~\ref{th:dual}. Therefore,
$Z^{-1}_{\delta_{\max}}\subseteq N\setminus Z_{\delta_{\max}}\subseteq
N\setminus Z_{\delta}$.

Let $T_{\delta}$ be the defining set of the dual code. Then
$T_{\delta}=(N\setminus Z_{\delta})^{-1} \supseteq
Z_{\delta_{\max}}$. Moreover $\{0\}\in N\setminus Z_{\delta}$ and
therefore $T_{\delta}$. Thus there are at least $\delta_{\max}$
consecutive roots in $T_{\delta}$. Thus the dual distance $d^\perp
\geq \delta_{\max}+1$.
\end{proof}

\begin{lemma}\label{th:hdualdist} 
Suppose that $C$ is a primitive, narrow-sense BCH code of length
$n=q^{2m}-1$ over $\F_{q^2}$ with designed distance $2\leq \delta\le
\delta_{\max}=q^{m+[\text{$m$ even}]}-1-(q^2-2)[m \textup{ even}])$, 
then the dual distance $d^\perp \geq
\delta_{\max} + 1$.
\end{lemma}
\begin{proof}
The proof is analogous to the one of Lemma~\ref{th:dualdist}; just
keep in mind that the defining set $Z_\delta$ is invariant under
multiplication by $q^2$ modulo $n$.
\end{proof}

\section{Families of Quantum Codes}\label{sec:qcodes}
We use the results of the previous sections to prove the existence of
quantum stabilizer codes.
\begin{theorem}\label{sh:euclid}
If $q$ is a power of a prime, and $m$ and $\delta$ are integers such that
$m\ge 2$ and $2\le \delta\le \delta_{\max}=q^{\lceil m/2\rceil}-1-(q-2)[m \text{ odd}]$,
then there exists a quantum stabilizer code $Q$ with parameters
$$[[q^m-1,q^m-1-2m\lceil(\delta-1)(1-1/q)\rceil, d_Q\ge \delta]]_q$$
that is pure up to $\delta_{\max}+1$. 
If $\B(n,q;\delta)$ has true minimum 
distance $d$, and $d\le \delta_{\max}$, then $Q$ is a pure quantum
code with minimum distance $d_Q=d$.
\end{theorem}
\begin{proof}
Theorem~\ref{th:bchdimension} and \ref{th:dual} imply that there
exists a classical BCH code with parameters
$[q^m-1,q^m-1-m\lceil(\delta-1)(1-1/q)\rceil,\ge \delta]_q$ which
contains its dual code. An $[n,k,d]_q$ code that contains its dual
code implies the existence of the quantum code with parameters
$[[n,2k-n,\ge d]]_q$ by the CSS construction, see~\cite{grassl04},
\cite{grassl99b}.  By Lemma~\ref{th:dualdist}, the dual distance
exceeds $\delta_{\max}$; the statement about the purity and minimum distance
is an immediate consequence. 
\end{proof}

\begin{theorem}\label{sh:hermite}
If $q$ is a power of a prime, $m$ is a positive integer, and $\delta$
is an integer in the range $2\le \delta \le q^{m}-1$, then there exists a
quantum code $Q$ with parameters
$$ [[q^{2m}-1, q^{2m}-1-2m\lceil(\delta-1)(1-1/q^2)\rceil , d_Q\ge
\delta]]_q$$ that is pure up to $\delta_{\max}+1$,
where $\delta_{\max}=q^{m+[\text{$m$ even}]}-1-(q^2-2)[m \textup{ even}])$.
If $\B(n,q^2;\delta)$ has
true minimum distance $d$, with $d\leq  \delta_{\max}$, then $Q$ is a pure
quantum code of minimum distance $d_Q=d$.
\end{theorem}
\begin{proof}
It follows from Theorems~\ref{th:bchdimension} and~\ref{th:hdual} that
there exists a primitive, narrow-sense $[q^{2m}-1,q^{2m}-1-m\lceil
(\delta-1)(1-1/q^2)\rceil,\ge\delta]_{q^2}$ BCH code that contains its
Hermitian dual code.  Recall that if a classical $[n,k,d]_{q^2}$ code
$C$ exists that contains its Hermitian dual code, then there exists an
$[[n,2k-n,\ge d]]_q$ quantum code that is pure up to $d$,
see~\cite{ashikhmin01}; this proves our claim. By
Lemma~\ref{th:hdualdist}, the Hermitian dual distance exceeds
$\delta_{\max}$, which implies the last statement of the claim. 
\end{proof}

\section{Conclusions}
We have investigated primitive, narrow-sense BCH codes in this note.
We were able to characterize when primitive, narrow-sense BCH codes
contain their Euclidean and Hermitian dual codes, and this allowed us
to derive two series of quantum stabilizer codes.  These results make
it possible to construct more families of quantum BCH codes as shown
by Cohen, Encheva, and Litsyn in \cite{cohen99}, since the BCH codes
are nested and are amenable to the Steane enlargement technique
\cite{steane99}. From a practical point of view, it is interesting
that efficient encoding and decoding algorithms are known for cyclic
binary quantum stabilizer codes, see~\cite{grassl00}.
 
\section*{Acknowledgment}
We thank Pascale Charpin for sending us a copy of~\cite{charpin98},
Simon Litsyn for sending us reference~\cite{cohen99}, and Neil Sloane for
very helpful discussions.

This research was supported by NSF CAREER award CCF~0347310,
NSF grant CCR 0218582 and a Texas A\&M TITF initiative. 

\enlargethispage{-15.3cm} % equalize page columns here
%\bibliographystyle{IEEEtranS}
%\bibliography{BCHreferences}

\end{document}